\documentclass[aps,prd,preprint,superscriptaddress,preprintnumbers,nofootinbib,amsmath,amssymb]{revtex4}

\usepackage{bookmark}
\usepackage{graphicx}
\usepackage{dcolumn}
\usepackage{bm}
\usepackage{subfigure}
\usepackage{amsmath}
\usepackage{bm}
\usepackage{slashed}
\usepackage{epsfig}
\usepackage{amsfonts}
\usepackage{color}
\usepackage{epstopdf}
\usepackage{enumitem}
\allowdisplaybreaks

\newcommand{\ud}{\mathrm{d}}

\newcommand{\re}{\mathop{\mathrm{Re}}\nolimits}

\begin{document}

\title{Relativistic corrections to gluon fragmentation into the $^3P_{J}^{[1,8]}$ states} 


\author{Zhi-Guo He}
 \email{zhiguo.he@buct.edu.cn}
\affiliation{Department of Physics and Electronics, School 
	of Mathematics and Physics, Beijing University of 
	Chemical Technology, Beijing 100029, China}
\affiliation{{II.} Institut f\"ur Theoretische Physik, 
	Universit\"at Hamburg,
	Luruper Chaussee 149, 22761 Hamburg, Germany}
\author{Bernd A. Kniehl}
 \email{kniehl@mail.desy.de}
\affiliation{{II.} Institut f\"ur Theoretische Physik, Universit\"at Hamburg,
	Luruper Chaussee 149, 22761 Hamburg, Germany}
\author{Peng Zhang}
 \email{pzhang@buct.edu.cn (corresponding author)}
\affiliation{Department of Physics and Electronics, School of Mathematics 
	and Physics, Beijing University of Chemical Technology, 
	Beijing 100029, China}
	
\date{\today}


\begin{abstract}
We compute relativistic corrections to the gluon fragmentation functions to
${}^3P_J^{[1,8]}$ Fock states of heavy quarkonium within non-relativistic QCD 
factorization framework. We find that, at $\mathcal{O}(v^2)$ sub-leading order, the 
$S$-$D$ mixing effect must be taken into account to absorb the infrared divergence 
of spin-triplet $P$-wave production within full QCD into the NRQCD long-distance 
matrix elements. Unlike the $S$-wave case, we find that the short-distance 
coefficients of the fragmentation functions at leading and sub-leading order are no 
longer proportional to each other. However, upon convolution with the gluon 
production cross section, their ratios are almost constant across the whole $p_T$ 
region. We find the relativistic corrections to be negative and substantial, which 
makes them a non-negligible ingredient in the study of $J/\psi$ production at the 
LHC.
	
\end{abstract}


\maketitle

\section{Introduction}
The non-relativistic QCD (NRQCD) factorization formalism~\cite{Bodwin:1994jh} that 
was constructed on top of the rigorous non-relativistic effective field 
theory~\cite{Caswell:1985ui} is now the most popular theoretical framework on the 
market to properly separate the perturbative and non-perturbative effects in heavy 
quarkonium production. In NRQCD factorization, the perturbative part describes the 
production of heavy quark pair that can be calculated through expansion in series of 
$\alpha_s$, and the non-perturbative part measures the hadronization of $Q\bar{Q}$ 
into the heavy meson by the supposedly universal long-distance matrix elements 
(LDMEs) whose sizes are governed by the velocity scaling rule. During the past years, 
great progress has been made to investigate the production mechanism of heavy 
quarkonium especially the $J/\psi$ meson in various collision processes such as 
$e^{+}e^{-}$, $\gamma\gamma$, $\gamma p$, $pp(\bar{p})$ (see 
Ref.~\cite{Lansberg:2019adr} and references therein for a recent review). However, 
several sets of LDMEs for $J/\psi$ production were 
reported~\cite{Ma:2010yw,Butenschoen:2011yh,Gong:2012ug,Bodwin:2014gia,
Brambilla:2024iqg}, which severely challenges the NRQCD hypothesis, and none of these 
LDME sets can explain the world data of $J/\psi$ yield and polarization. In many 
cases, theoretical calculation exhibited that the next-to-leading order (NLO) QCD 
corrections are huge and may even become orders of magnitude larger than leading 
order (LO) results. Therefore, it is not yet conclusive whether the inconsistency 
among the LDMEs originates from the poor convergence of perturbative calculation or 
the incapability of NRQCD factorization itself.    

The situation becomes relatively clear for $J/\psi$ hadroproduction in large transverse 
momentum ($p_T$) region, where the fragmentation mechanism plays the major 
role~\cite{Braaten:1993rw,Braaten:1993mp}. Based on QCD factorization, it was 
proposed that the perturbative calculation in such region should be re-organized 
according to the new $p_T$ power counting 
rule~\cite{Kang:2011zza,Kang:2011mg,Kang:2014tta} about the partonic cross section. 
At leading power (LP) the partonic cross section behaves as $1/p_T^4$ due to single 
parton fragmentation contribution, and at next-to leading power (NLP) it behaves as 
$1/p_T^6$ due to double parton fragmentation contribution. By combining pQCD and NRQCD 
factorization, the fragmentation function (FF) can be further factorized into the sum 
of short-distance coefficients (SDCs) and the corresponding NRQCD LDMEs.

For prompt $J/\psi$ production the most relevant Fock states include CS $^3S_1^{[1]}$, 
$^3P_{J}^{[1]}$ with $J=1,2$ and CO $^3S_1^{[8]}$,$^1S_0^{[8]}$ and $^3P_J^{[8]}$ with 
$J=0,1,2$. Except for the gluon fragment into $^3S_1^{[1]}$ channel, the analytical 
results of the SDCs for all the other single gluon or heavy quark FFs at QCD LO were 
obtained about 3 decades ago~ 
\cite{Braaten:1993mp,Braaten:1993rw,Braaten:1994kd,Cho:1994gb,Ma:1995vi,Braaten:1995cj,
Beneke:1995yb,Braaten:1996rp}, which greatly helped to understand the prompt $J/\psi$ and 
$\psi^{\prime}$ hadroproduction at the Tevatron. The SDCs for single and double parton 
FF up to $\mathcal{O}(\alpha_{s}^2)$ were summarized in Ref.~\cite{Ma:2013yla,Ma:2015yka}, 
and the analytical form of SDCs for gluon fragment into $c\bar{c}(^3S_1^{[1]})$ were 
firstly given by Zhang et al.~\cite{Zhang:2017xoj}. To improve the precision of 
theoretical predictions the NLO QCD corrections to gluon fragment into 
$c\bar{c}(^3S_1^{[8]})$~\cite{Braaten:2000pc},
$c\bar{c}(^1S_0^{[8]})$~\cite{Artoisenet:2018dbs,Feng:2018ulg,Zhang:2018mlo}, and 
$c\bar{c}(^3P_J^{[1,8]})$~\cite{Zhang:2020atv} were computed by different groups 
independently, as well as the charm quark fragmenting into 
$c\bar{c}(^3S_1^{[1]})$~\cite{Sepahvand:2017gup,Zheng:2019dfk} and $c\bar{c}(^1S_0^{[8]})$ ~\cite{Feng:2021uct}.

Besides the NLO QCD corrections, it has been found that higher order relativistic 
corrections are also considerable for $J/\psi$ associated production with 
$\eta_c$~\cite{He:2007te} or light hadrons~\cite{He:2009uf,Jia:2009np} in $e^{+}e^{-}$ 
annihilation, for $J/\psi$ yield~\cite{Xu:2012am,He:2014sga} and 
polarization~\cite{He:2015gla} in photo- and hadroproduction, and for double $J/\psi$ 
hadroproduction near threshold~\cite{He:2024ugx}. Therefore, it would be of great interest
to go on to explore the higher order relativistic effects in the FFs. The FF of gluon and 
heavy quark fragment into the CS $^3S_1^{[1]}$ were already known up to $\mathcal{O}(v^4)$~\cite{Bodwin:2003wh,Bodwin:2012xc,Sang:2009zz,Cui:2025wjq}. 
However, for the CO state, only the relativistic corrections of gluon fragment into the 
S-wave $^3S_1^{[8]}$ and $^1S_0^{[8]}$ were computed and in which the SDCs at the $v^2$ 
sub-leading were found to be proportional to those at LO by a common trivial factor 
$-\frac{11}{6}m_c^2$~\cite{Bodwin:2003wh,Gao:2016ihc} suggesting the NLO $v^2$ 
corrections to the fragmentation function to be as important as the NLO QCD corrections. 

Now the only barrier on road from a complete analysis of the LP contribution to prompt 
$J/\psi$ hadroproduction up to $v^2$ sub-leading order accuracy is the missing of 
relativistic corrections to gluon fragment into the CS and CO $P$-wave states 
$^3P_J^{[1,8]}$. Theoretically it is also curious to know if the SDCs of FF in P-wave 
cases at $\mathcal{O}(v^2)$ would be proportional to those at $\mathcal{O}(v^2)$ as the 
S-wave cases. All of the above motivate us to calculate the relativistic corrections to 
gluon fragment into the CS and CO $P$-wave states $^3P_J^{[1,8]}$ completing the last 
$v^2$ sub-leading order pieces of gluon FFs for prompt $J/\psi$ hadroproduction. In full 
QCD calculation of P-wave Fock states production at $v^2$ LO, there are infrared 
divergences that are absorbed into the NLO QCD corrections of the $S$-wave NRQCD 
LDMEs~\cite{Braaten:1994kd,Ma:2013yla}. At $v^2$ sub-leading order, the structures of the 
infrared divergences will be more complicated and an additional $S$-$D$ mixing 
contribution will be needed to remove all of them. Therefore, the LO $S$-$D$ term of the 
gluon FF will be given for the first time as well. 

The remainder of this paper is organized as follows. In Sec.~\ref{sec:framework}, we will 
introduce the definition of gluon FF for heavy quarkonium production and its conjunction 
with NRQCD factorization. In Sec.~\ref{sec:calculation}, we will describe the details to 
calculate the SDCs of relativistic corrections to $^3P_J^{[1,8]}$ through matching 
between full QCD and NRQCD, and render the final infrared finite results. In  
Sec.~\ref{sec:numerical}, we will discuss comprehensively about the potential influence 
of the NLO relativistic corrections, and the conclusion will be contained in the end.

\section{Heavy quarkonium fragmentation function}\label{sec:framework}
The FF for heavy quarkonium production can either be calculated directly from the Feynman 
diagram of gluon or heavy quark fragmenting into heavy quarkonium~\cite{Braaten:1993rw} or in a formal way following the rigorous operator definition introduced by Collins and 
Soper~ \cite{Collins:1981uw}. We perform independent calculation in both ways, and get 
the same results. It is straightforward to compute the gluon fragmenting Feynman 
diagrams, thus here we only construe the formal approach that we used. 

In Collins-Soper definition, it is more convenient to re-write a four-vector $V$ in 
light-cone coordinate as $V=(V^{+},V^{-},\mathbf{V}_{\perp})$ with 
$V^{+}=(V^{0}+V^{3})/\sqrt{2}$, $V^{-}=(V^{0}-V^{3})/\sqrt{2}$, and 
$\mathbf{V}_{\perp}=(0,V_1,V_2,0)$. The gauge invariant definition of FF for $g\to H$ can 
then be expressed as~\cite{Collins:1981uw}
\begin{align}\label{eq:defFF}
D_{g \rightarrow H}(z,\mu_0)= \frac{-g_{\mu\nu}z^{D-3}}{2 \pi P_g^{+}(N_{c}^{2}-1)(D-2)} \int_{-\infty}^{+\infty}\mathrm{d}x^{-} e^{-i P_g^{+} x^{-}} \nonumber \\ 
\times \langle 0 | G_{c}^{+\mu}(0) \mathcal{E}^{\dag}(0,0,\boldsymbol{0}_{\perp})_{cb} \mathcal{P}_{H(P)} \mathcal{E}(0,x^{-},\boldsymbol{0}_{\perp})_{ba} G_{a}^{+\nu}(0,x^{-},\boldsymbol{0}_{\perp}) | 0 \rangle \, ,
\end{align}
where $P_g$ and $P$ are the momenta of the initial fragmenting gluon and final state 
heavy quarkonium $H$, respectively, $z=P^+/P_g^+$ indicating the momentum fraction of the 
initial gluon in ``+" direction, $G_c^{\mu\nu}$ is the gluon field-strength operator, and 
$\mathcal{P}_{H(P)}$ is the projection operator onto a state of heavy quarkonium $H$ with 
anything, $X$, formulated as 
\begin{equation}\label{eq:projectH}
	\mathcal{P}_{H(P)} = \sum_X |H(P)+X \rangle \langle H(P)+X|\,.
\end{equation}
Since there are infrared divergence in the full QCD calculation, we adopt the dimensional 
regularization (DR) scheme to regularize the divergence with $D=4-2\epsilon$ as the 
dimension of space-time. We evaluate the FF in such a frame that the transverse 
components of the $H$ four-momentum is 0, i.e. 
$P = (z P_g^+, M^2/(2 zP_g^+),\boldsymbol{0}_{\perp})$ with $P^2=2P^+P^-=M^2$. 
The gauge link $\mathcal{E}(x^{-})$, which ensures the gauge invariance of FF, entails a 
path-ordered exponential of gluon field $A^{\mu}$ along lightlike path,
\begin{equation}
  \mathcal{E}(0,x^{-},\boldsymbol{0}_{\perp})_{ba}= \mathcal{P} \, \text{exp} \left[i g_s \int_{x^{-}}^{\infty}\mathrm{d}z^{-} A^{+}(0,z^{-},\boldsymbol{0}_{\perp}) \right]_{ba} \, ,
\end{equation}
where $g_s = \sqrt{4 \pi \alpha_s}$ denotes the QCD coupling constant.

Under the hypothesis that NRQCD factorization also works for the FFs, then up to relative 
$v^2$ order $D_{g \rightarrow H}(z)$ can be further factorized as
\begin{eqnarray}
D_{g \rightarrow H}(z)=\sum_n d^{(0)}[g\to c\bar{c}(n)] \langle\mathcal{O}^{H}(n)\rangle + d^{(2)}[g\to c\bar{c}(n)] \frac{\langle\mathcal{P}^{H}(n)\rangle}{m_c^2},
\end{eqnarray}
where $\mathcal{O}^{H}(n)$ is the four-fermion operators depicting the transition of Fock 
state $n$ into heavy quarkonium $H$ at $v^2$ LO, $\mathcal{P}^{H}(n)$ is its $v^2$ 
correction, and $d_n^{(0)}(z)$ and $d_n^{(2)}(z)$ are the corresponding SDCs. The 
definitions of the $\mathcal{O}$ operators for $n=$$^3P_J^{[1,8]}$ can be found in 
Ref.~\cite{Bodwin:1994jh}, and we define the corresponding $\mathcal{P}$ as
\begin{eqnarray}
&&\mathcal{P}^{J/\psi}(^{3}P_{J}^{[8]})=\chi^{\dagger}\boldsymbol{\sigma}^{i}
\left(-\frac{i}{2}\overleftrightarrow{\boldsymbol{D}^{j}}\right)T^{a}\psi
(a^{\dagger}_{J/\psi}a_{J/\psi})\psi^{\dagger}\boldsymbol{\sigma}^{i}T^{a}
\left(-\frac{i}{2}\overleftrightarrow{\boldsymbol{D}^{j}}\right)
\left(-\frac{i}{2}
\overleftrightarrow{\boldsymbol{D}}\right)^{2}\chi+{\rm H.c.},\nonumber\\
&&\mathcal{P}^{\chi_{c0}}(^{3}P_{0}^{[1]})=\frac{1}{6N_c}\chi^{\dagger}
\left(-\frac{i}{2}\overleftrightarrow{\boldsymbol{D}}\cdot\boldsymbol{\sigma}\right)\psi
(a^{\dagger}_{\chi_{c0}}a_{\chi_{c0}})\psi^{\dagger}\left(-\frac{i}{2}
\overleftrightarrow{\boldsymbol{D}}\cdot\boldsymbol{\sigma}\right)
\left(-\frac{i}{2}\overleftrightarrow{\boldsymbol{D}}\right)^{2}\chi+{\rm H.c.},
\nonumber\\
&&\mathcal{P}^{\chi_{c1}}(^{3}P_{1}^{[1]})=\frac{1}{4N_c}\chi^{\dagger}
\left(-\frac{i}{2}\overleftrightarrow{\boldsymbol{D}}\times\boldsymbol{\sigma}\right)\psi
(a^{\dagger}_{\chi_{c1}}a_{\chi_{c1}})\psi^{\dagger}\left(-\frac{i}{2}
\overleftrightarrow{\boldsymbol{D}}\times\boldsymbol{\sigma}\right)
\left(-\frac{i}{2}
\overleftrightarrow{\boldsymbol{D}}\right)^{2}\chi+{\rm H.c.},
\nonumber\\
&&\mathcal{P}^{\chi_{c2}}(^{3}P_{2}^{[1]})=\frac{1}{4N_c}\chi^{\dagger}
\left(-\frac{i}{2}\overleftrightarrow{\boldsymbol{D}}^{(i}\boldsymbol{\sigma}^{j)}\right)
\psi(a^{\dagger}_{\chi_{c2}}a_{\chi_{c2}})\psi^{\dagger}\left(-\frac{i}{2}
\overleftrightarrow{\boldsymbol{D}}^{(i}\boldsymbol{\sigma}^{j)}\right)
\left(-\frac{i}{2}\overleftrightarrow{\boldsymbol{D}}\right)^{2}\chi+{\rm H.c.},
\end{eqnarray}
where
$\overleftrightarrow{\boldsymbol{D}}^{(i}\boldsymbol{\sigma}^{j)}
=(\overleftrightarrow{\boldsymbol{D}}^{i}\boldsymbol{\sigma}^{j}
+\overleftrightarrow{\boldsymbol{D}}^{j}\boldsymbol{\sigma}^{i})/2
-\overleftrightarrow{\boldsymbol{D}}\cdot\boldsymbol{\sigma}\delta^{ij}/3$ and
$\boldsymbol{K}^{ij}
=(-i/2)^2(\overleftrightarrow{\boldsymbol{D}}^{i}
\overleftrightarrow{\boldsymbol{D}}^{j}
-\overleftrightarrow{\boldsymbol{D}}^{2}\delta^{ij}/3)$. 
Note that there is a $1/2/N_c$ normalization factor difference between our CS operators 
and those in Ref.~\cite{Bodwin:1994jh}.

To cancel the infrared divergence, we also need 2 more additional $\mathcal{P}$ operators
\begin{eqnarray}
&&\mathcal{P}^{H}(^{3}S_{1}^{[8]})=\chi^{\dagger}\boldsymbol{\sigma}^{i}T^{a}\psi
(a^{\dagger}_{H}a_{H})\psi^{\dagger}\boldsymbol{\sigma}^{i}T^{a}\left(-\frac{i}{2}
\overleftrightarrow{\boldsymbol{D}}\right)^{2}\chi+{\rm H.c.},
\nonumber\\
&&\mathcal{P}^{H}(^{3}S_{1}^{[8]},^{3}D_{1}^{[8]})=\sqrt{\frac{3}{5}}\chi^{\dagger}
\sigma^{i}T^{a}\psi (a^{\dagger}_{H}a_{H})\psi^{\dagger}\boldsymbol{\sigma}^{j}
\boldsymbol{K}^{ij}T^{a}\chi+{\rm H.c.},
\label{eq:pop}
\end{eqnarray}
Our definition of $S$-$D$ mixing operator $\mathcal{P}^{H} (^{3}S_{1}^{[8]},^{3}D_{1}^{[8]})$ is different from that in Refs.~\cite{Bodwin:1994jh}, in which a linear combination of $\mathcal{P}^{H}(^{3}S_{1}^{[8]})$ and
$\mathcal{P}^{H}(^{3}S_{1}^{[8]},^{3}D_{1}^{[8]})$ in Eq.~(\ref{eq:pop}) is used instead.
At order $v^2$, the heavy-quark spin symmetries among the $\mathcal{O}^H(n)$ operators break down, but they still hold among the $\mathcal{P}^H(n)$ operators, yielding the relationships
\begin{eqnarray}
	\langle\mathcal{P}^{J/\psi}(^{3}P_{0}^{[8]})\rangle
	&=&\frac{1}{2J+1}\langle\mathcal{P}^{J/\psi}(^{3}P_{J}^{[8]})\rangle,
	\nonumber\\
	\langle\mathcal{P}^{\chi_{c0}}(^{3}S_{1}^{[8]})\rangle
	&=&\frac{1}{2J+1}\langle\mathcal{P}^{\chi_{cJ}}(^{3}S_{1}^{[8]})\rangle,
	\nonumber\\
	\langle\mathcal{P}^{\chi_{c0}}(^{3}P_{0}^{[1]})\rangle
	&=&\frac{1}{2J+1}\langle\mathcal{P}^{\chi_{cJ}}(^{3}P_{J}^{[1]})\rangle,
	\nonumber\\
	\langle\mathcal{P}^{\chi_{c0}}(^{3}S_{1}^{[8]},^{3}D_{1}^{[8]})\rangle
	&=&-\frac{2}{3}\langle\mathcal{P}^{\chi_{c1}}(^{3}S_{1}^{[8]},^{3}D_{1}^{[8]})\rangle
	=2\langle\mathcal{P}^{\chi_{c2}}(^{3}S_{1}^{[8]},^{3}D_{1}^{[8]})\rangle.
\end{eqnarray}

In NRQCD factorizatin, the SDCs describe physics above the NRQCD factorization scale 
$\mu_{\Lambda}$, therefore $d_{g\to c\bar{c}(n)}^{(0)}(z)$ and 
$d_{g\to c\bar{c}(n)}^{(2)}(z)$ can be determined through 
matching between full QCD and NRQCD calculation of the FF, $D[g\to c\bar{c}(n)]$, for free 
$c\bar{c}$ production in configuration $n$. 

\section{Perturbative calculation}\label{sec:calculation}
\subsection{Full QCD calculation}
\begin{figure}[htb!]
	\begin{center}
		\includegraphics[width=0.25\textwidth]{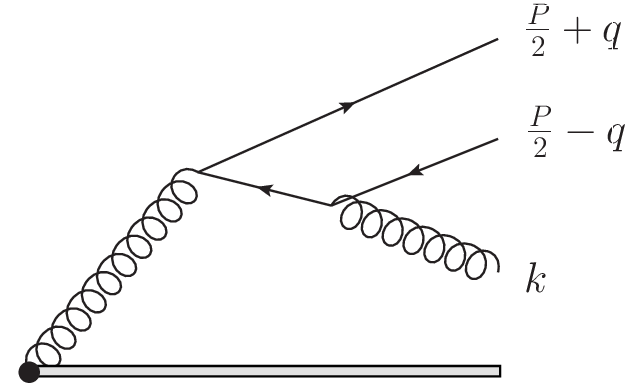} 
		\caption{The representative Feynman diagrams for gluon fragmenting into $c\bar{c}(^3P_J^{[1,8]})$ at $\alpha_s$ LO. \label{fig:lofd}}
	\end{center}
\end{figure}

At LO in $\alpha_s$, the gluon fragment into $c\bar{c}$ in $^3P_J^{[1,8]}$ through 
$g\to c\bar{c}(^3P_J^{[1,8]})+g$ process, the representative Feynman diagram of which is 
shown in Fig.~\ref{fig:lofd}. The amplitude for production of $c\bar{c}$ in a specific 
configuration $n=$$^{2S+1}L_{J}^{[1,8]}$ can be calculated directly with the help of spin 
projection method. The color projector onto the CS or CO state $a$ are given 
by~\cite{Petrelli:1997ge}
\begin{equation}
\Lambda_1=\langle3,k;\bar{3},l|1\rangle=\delta_{kl}/\sqrt{N_c},\;
\Lambda_8=\langle3,k;\bar{3},l|8,a\rangle=\sqrt{2}\,T^{a}_{kl} 
\end{equation}
where $N_c=3$, and $T^a$ are the generators of SU(3) color group in its fundamental 
representation. While the projector holding to all orders in $v$ to project the product 
of Dirac spinors $v(\bar{p})\bar{u}(p)$ onto spin-triplet state with spin vector 
$\epsilon_S$ is written as~\cite{Bodwin:2002cfe}
\begin{align}
	\Lambda(\epsilon^{\ast}_S)
	= \frac{1} {\sqrt{2 E} (E + m_c)}
	(\slashed{\overline{p}} - m_c)
	 \slashed{\epsilon}_S^{\ast}
	\frac{2 E + \slashed{P}}{4 E}
	(\slashed{p} + m_c)	\, ,
\end{align}
where $p$ and $\bar{p}$ are momenta of heavy quark and anti-quark, and  
$E=\sqrt{m_c^2+\boldsymbol{q}^2}$. Their relations with the total momentum $P$ and the 
relative momentum $q$ are:
\begin{equation}
    p=\frac{P}{2}+q, \quad \bar{p}=\frac{P}{2}-q, \quad P^2=M^2=4E^2,
\end{equation}
and in $c\bar{c}$ rest frame, $P=(2E,\boldsymbol{0})$, $q=(0,\boldsymbol{q})$. 

Then the projected amplitude $\mathcal{M}$ on spin-triplet and CS with c=1 or CO with c=8 can be expressed as
\begin{equation}
	\mathcal{M}(q) =\text{Tr}[\mathcal{\overline{M}} (\Lambda(\epsilon^{\ast}_S) \otimes \Lambda_c)] \, ,
\end{equation}
where $\overline{\mathcal{M}}$ is the spinor truncated standard Feynman amplitude, and the
trace is understood to take over both Dirac and color indices.

To get the $v^2$ sub-leading order corrections, we need to expand not only the amplitude 
but also phase space integration in series of $q^2$. The phase space measure of fragmentation function can be expressed as follows~\cite{Zhang:2018mlo}:
\begin{equation} \label{eq:phase}
	\mathrm{d} \Phi =
	\frac {P^{+}}{z^{2}}
	\delta \left( \frac{1-z}{z} P^{+} - k^{+} \right)
	\frac{\mathrm{d}^{D} k}{(2\pi)^{D-1}} \delta_+ (k^{2}) \,,
\end{equation}
where $k$ is the momentum of gluon in the final state, and $\delta_+(k^2)$ indicates that only the positive energy solution of the on-shell condition $k^2=0$ is considered. We introduce the dimensionless variables 
\begin{equation} \label{eq:dless}
	\hat{P} = \frac {P}{M} \, , \quad
	\hat{k} = \frac {k}{M} \, .
\end{equation}
to factor $M$ out, in such a way $\mathrm{d} \Phi$ can be re-written as
\begin{equation}
	\ud \Phi =M^{D-2}\frac {\hat{P}^{+}}{z^{2}}
	\delta \left( \frac{1-z}{z} \hat{P}^{+} - \hat{k}^{+} \right)
	\frac{\mathrm{d}^{D} \hat{k}}{(2\pi)^{D-1}} \delta_+ (\hat{k}^{2})= M^{D-2} \ud \hat{\Phi} .
\end{equation}
Now the re-scaled phase space $\ud \hat{\Phi}$ does not depend on $q$ anymore. 
Absorbing the factor $M^{D-2}$ into the squared amplitude, we thus only need to expand 
the re-scaled amplitude
\begin{equation}
	\hat{\mathcal{M}}(\hat{P},\hat{k},M,q) =  M^{(D-2)/2} \mathcal{M}(M \hat{P},M \hat{k},q) \, .
\end{equation}
in series of the relative momentum $q$,
\begin{align}\label{eq:amplitudeExp}
	\hat{\mathcal{M}}(q) = \hat{\mathcal{M}}(0)+ q_{\beta} \hat{\mathcal{M}}^{\beta}(0)
	+ q_{\beta_1} q_{\beta_2} \hat{\mathcal{M}}^{\beta_1\beta_2}(0) 
	+ q_{\beta_1} q_{\beta_2} q_{\beta_3}  \hat{\mathcal{M}}^{\beta_1\beta_2\beta_3}(0) +\cdots \, .
\end{align}
where
\begin{equation}
	\hat{\mathcal{M}}^{\beta_1\cdots \beta_N} (0)= \frac{1}{N!} 
	\frac{\partial^N \hat{\mathcal{M}}(q)}{\partial q_{\beta_1} 
		\cdots \partial q_{\beta_N}} \Big|_{q\to0} \, ,
\end{equation}
and recall that $M=2\sqrt{m_c^2-q^2}$. For $P$-wave states only terms with odd power in 
$q$, and it is other way around for $S$-wave and $D$-wave states. To calculate the 
contribution beyond $v^2$ LO, we need to further decompose the higher-rank tensor of $q$ 
product into the irreducible ones for $S$- and $P$-wave cases as
	\begin{align}
		q_{\beta_1} q_{\beta_2}  \to \frac{|\boldsymbol{q}|^2}{D-1} \Pi_{\beta_1\beta_2}, \quad
		q_{\beta_1} q_{\beta_2} q_{\beta_3}  \to -\frac{|\boldsymbol{q}|^2}{D+1} q^\beta \Pi_{\beta_1\beta_2\beta_3\beta} \, .
	\end{align}	
where  
\begin{subequations}
	\begin{align}
		\Pi_{\beta_1\beta_2} &= -g_{\beta_1\beta_2} + \frac{\hat{P}_{\beta_1}\hat{P}_{\beta_2}}{\hat{P}^2} \, , \\
		\Pi_{\beta_1\beta_2\beta_3\beta_4} &= \Pi_{\beta_1\beta_2} \Pi_{\beta_3\beta_4} + \Pi_{\beta_1\beta_3} \Pi_{\beta_2\beta_4} + \Pi_{\beta_1\beta_4} \Pi_{\beta_2\beta_3}   \, .
	\end{align}	
\end{subequations}

The FF function for $g\to c\bar{c} (^3P_J^{[1,8]})+g$ production in full QCD 
calculation then can be written as
\begin{align}\label{eq:fffullqcd} 
D[g\to c\bar{c}(^3P_J^{[1,8]})]&=\tilde{d}^{(0)}[g\to c\bar{c}(^3P_J^{[1,8]})] \langle0|\mathcal{O}(^3P_J^{[1,8]})|0\rangle|_{\mathrm{QCD}}\nonumber\\
&+\tilde{d}^{(2)}[g\to c\bar{c}(^3P_J^{[1,8]})] \frac{\langle0|\mathcal{P}(^3P_J^{[1,8]})|0\rangle|_{\mathrm{QCD}}}{m_c^2}\, , 
\end{align}
where the short-distance parts in full QCD calculation are
\begin{subequations}
\begin{eqnarray}
\tilde{d}^{(0)}[g\to c\bar{c}(^3P_J^{[1,8]})]=\frac{z^{D-2}}{(N_c^2-1)(D-2)}\int \overline{\sum_{S_z,L_z,J_z}}
	\left|\langle1,L_z;1,S_z|J,J_z\rangle
	\epsilon^{\ast\beta}_{L_z}\hat{\mathcal{M}}_{\beta}(0)\right|^2 d\hat{\Phi}
\end{eqnarray}
\begin{eqnarray}		
\frac{\tilde{d}^{(2)}[g\to c\bar{c}(^3P_J^{[1,8]})]}{m_c^2}=\frac{z^{D-2}}{(N_c^2-1)(D-2)}\int
{}\frac{-2}{D+1}\overline{\sum_{S_z,L_z,J_z}}\langle1,L_z;1,S_z|J,J_z\rangle
\nonumber\\
\times\overline{\sum_{S^{\prime}_z,L^{\prime}_z,J^{\prime}_z}}
\langle1,L^{\prime}_z;1,S^{\prime}_z|J,J^{\prime}_z\rangle\Pi_\beta^{\phantom{\beta}\beta_1\beta_2\beta_3}
\re\left[\epsilon^{\ast\beta^{\prime}}_{L_z}\epsilon^{\beta}_{L^{\prime}_z}
\hat{\mathcal{M}}_{\beta^{\prime}}(0)\hat{\mathcal{M}}^*_{\beta_1\beta_2\beta_3}(0)\right]
 d\hat{\Phi}
\end{eqnarray}
\end{subequations}
In the above equation, $\epsilon_{L_z}$ is the polarization four-vector for $L=P$, and 
the symbol $\overline{\sum}$ means summing over the polarization and color of all the 
initial and final states and then divided by the total degree of freedom of 
$^3P_J^{[1,8]}$ state in color and polarization space.  

We notice that the phase space integrals can be written as the sum of some basic ones in 
the following form
\begin{equation}\label{eq:mi}
	I_n = \int \ud \hat{\Phi} \, \frac{1}{(\hat{k} \cdot \hat{P}+a)^n}. \, 
\end{equation}
which can be integrated out analytically yielding
\begin{align}
	I_n =  \frac{ \Gamma (n-1+\epsilon) }{8 \pi ^2 \Gamma (n) z^2 }
	\left( \frac{\pi \mu_r^2}{m_c^2} \right)^{\epsilon}
	\left(\frac{z}{2 (1-z)}\right)^{\epsilon } \left(\frac{1-z}{2 z}+a\right)^{-(n-1+\epsilon)}  
\end{align}
where $\mu_r$ is the renormalization scale in dimensional regularization. The infrared 
divergences originate from soft gluon emission in the end $z=1$ point region, and they 
can be isolated by using the identity  
\begin{align}
	(1-z)^{-1-2\epsilon} =  -\frac{\delta(1-z)}{2 \epsilon }+\frac{1}{ [1-z]_+} 
	-2 \epsilon  \left[ \frac{\ln(1-z)  }{1-z}\right]_+ + \mathcal{O} (\epsilon^2) \,,
\end{align}
where the ``+" function is defined through the integration
\begin{eqnarray}
\int^{1}_0 g(z) f(z)_{+}dz\equiv\int^{1}_0 (g(z)-g(1)) f(z)dz.
\end{eqnarray}
Then up to $\mathcal{O}(\epsilon^{0})$ the analytical results of CS 
$\tilde{d}^{(0)}[g\to c\bar{c}(^3P_J^{[1]})]$ and $\tilde{d}^{(2)}[g\to c\bar{c}(^3P_J^{[1]})]$ are

\begin{subequations}\label{qcd_p_ff}
	\begin{align}
		\tilde{d}^{(0)}[g\to c\bar{c}(^3P_0^{[1]})] = & f(\epsilon)
		\Big( -\frac{2}{9 \epsilon} \delta(1-z)-\frac{1}{27} \delta(1-z) 
		+\frac{4}{9[1-z]_+}   \nonumber \\
		&+\frac{-26 z^2+85 z-8}{18}+ (5-3 z) \ln(1-z) \Big) \,, 
	\end{align}
	\begin{align}
		\tilde{d}^{(2)}[g\to c\bar{c}(^3P_J^{[1]})] = & f(\epsilon)
		\Big(\frac{31 \delta(1-z)}{45 \epsilon }+\frac{407}{1350} \delta(1-z)-\frac{62}{45 [1-z]_+} \nonumber \\
		&+\frac{710 z^2-2419 z+248}{180} +\frac{255 z-433}{30} \ln(1-z)
		\Big) \,.
	\end{align}
	\begin{align}
		\tilde{d}^{(0)}[g\to c\bar{c}(^3P_1^{[1]})] = & f(\epsilon)
		\Big( -\frac{2}{9 \epsilon} \delta(1-z)-\frac{5}{54} \delta(1-z) 
		+\frac{4}{9[1-z]_+} +\frac{-4 z^2-z-4}{9}  \Big),
	\end{align}
	\begin{align}
		\tilde{d}^{(2)}[g\to c\bar{c}(^3P_1^{[1]})] = & f(\epsilon) 
		\Big(\frac{7 \delta(1-z)}{15 \epsilon }+\frac{343}{900} \delta(1-z) 
		-\frac{14}{15 [1-z]_+}+\frac{7(4 z^2+z+4)}{30}\Big) \,,
	\end{align}
	\begin{align}
		\tilde{d}^{(0)}[g\to c\bar{c}(^3P_2^{[1]})] = & f(\epsilon)
		\Big( -\frac{2}{9 \epsilon} \delta(1-z)-\frac{19}{270} \delta(1-z)+\frac{4}{9[1-z]_+} \nonumber \\
		& +\frac{-4 z^2+11 z-4}{9} -\frac{4(z-2)}{5} \ln(1-z) \Big)
	\end{align}
	\begin{align}	
		\tilde{d}^{(2)}[g\to c\bar{c}(^3P_2^{[1]})]= & f(\epsilon) 
		\Big(\frac{5 \delta(1-z)}{9 \epsilon }+\frac{43}{108} \delta(1-z)-\frac{10}{9 [1-z]_+} \nonumber \\
		&+\frac{476 z^2-1375 z+500}{450}+\frac{2(75 z-152)}{75} \ln(1-z) 
		\Big) \,,
	\end{align}
\end{subequations}
The over all factor $f(\epsilon)$ is 
$f(\epsilon) = \frac{ \, \alpha_s^2 }{ N_c m_c^5 }\left( \frac{\pi \mu_r^2}{m_c^2} \right)^{\epsilon} \Gamma(1+\epsilon)$. 
The $\tilde{d}^{(0,2)}[g\to c\bar{c}(^3P_J^{[8]})]$ for CO states are only different from 
the corresponding $\tilde{d}^{(0,2)}[g\to c\bar{c}(^3P_J^{[1]})]$ by a color factor 
$(N_c^2-4)/(2(N_c^2-1))$. Next we will show how to absorb these infrared divergences into 
NRQCD LDMEs leading to finite SDCs. 

\subsection{NRQCD calculation and Matching}
According to NRQCD factorization, up to the $v^2$ sub-LO the FFs of 
$g\to c\bar{c}(^3P_J^{[1,8]})$ can be written as
\begin{eqnarray}\label{nrqcd_p_ff}
&&D[g\to c\bar{c}(^3P_J^{[1,8]})]=
\nonumber\\
&&d^{(0)}[g\to c\bar{c}(^3P_J^{[1,8]})] \langle0|\mathcal{O}(^3P_J^{[1,8]})|0\rangle|_{\mathrm{NRQCD}}
+d^{(0)}[g\to c\bar{c}(^3S_1^{[8]})] \langle0|\mathcal{O}(^3S_1^{[8]})|0\rangle|_{\mathrm{NRQCD}}
\nonumber\\
&&+d^{(2)}[g\to c\bar{c}(^3P_J^{[1,8]})] \frac{\langle0|\mathcal{P}(^3P_J^{[1,8]})|0\rangle|_{\mathrm{NRQCD}}}{m_c^2}
+d^{(2)}[g\to c\bar{c}(^3S_1^{[8]})] \frac{\langle0|\mathcal{P}(^3S_1^{[8]})|0\rangle|_{\mathrm{NRQCD}}}{m_c^2}
\nonumber\\
&&+d^{(2)}[g\to c\bar{c}(^3S_1^{[8]},^3D_1^{[8]})] \frac{\langle0|\mathcal{P}(^3S_1^{[8]},^3D_1^{[8]})|0\rangle|_{\mathrm{NRQCD}}}{m_c^2}
\end{eqnarray}
To get $d^{(0,2)}[g\to c\bar{c}(^3P_J^{[1,8]})]$ via matching, we need first to know the 
SDCs $d^{(0,2)}[g\to c\bar{c}(^3S_1^{[8]})]$ and 
$d^{(2)}[g\to c\bar{c}(^3S_1^{[8]},^3D_1^{[8]})]$ at QCD LO and the corresponding LDMEs 
at $\mathcal{O}(\alpha_s)$, except for $\langle0|\mathcal{O}^{c\bar{c}(^3P_J^{[1,8]})}(^3S_1^{[8]})|0\rangle|_{\mathrm{NRQCD}}$
of which the $\mathcal{O}(\alpha_s v^2)$ terms also contribute. 

$d^{(0,2)}[g\to c\bar{c}(^3S_1^{[8]})]$ and 
$d^{(2)}[g\to c\bar{c}(^3S_1^{[8]},^3D_1^{[8]})]$ can be computed using the spin project 
method directly, since there are no infrared divergences. Their results in $D$ dimension are
\begin{subequations}\label{swave_sdc}
\begin{align}
d^{(0)}[g\to c\bar{c}(^3S_1^{[8]})]&=\frac{z^{D-2}}{(N_c^2-1)(D-2)}\int \overline{\sum_{S_z}}
\left|\hat{\mathcal{M}}(0)\right|^2 d\hat{\Phi}
\nonumber\\
&=\frac{\pi  \alpha_s}{ (D-1) (N_c^2-1) m_c^3}\delta (1-z),
\end{align}	
\begin{align}
\frac{d^{(2)}[g\to c\bar{c}(^3S_1^{[8]})]}{m_c^2}&=\frac{z^{D-2}}{(N_c^2-1)(D-2)}\int\frac{2}{D-1} \overline{\sum_{S_z}}
\re\left[\Pi^{\beta_1\beta_2}\hat{\mathcal{M}}_{\beta_1\beta_2}(0)\hat{\mathcal{M}}^{\ast}(0)\right]^2 d\hat{\Phi}\nonumber\\
&=-\frac{(3D-1)\pi \alpha_s}{ 2(D-1)^2 (N_c^2-1) m_c^5}\delta (1-z),
\end{align}
\begin{align}
&\frac{d^{(2)}[g\to c\bar{c}(^3S_1^{[8]},^3D_1^{[8]})]}{m_c^2}=\frac{z^{D-2}}{(N_c^2-1)(D-2)}\int
\overline{\sum_{S_z,L_z,J_z}}\langle2,L_z;1,S_z|1,J_z\rangle
\nonumber\\
& 2\re\left[ \epsilon_{L_z}^{\ast\beta_1\beta_2}\mathcal{M}_{\beta_1\beta_2}(0)
\mathcal{M}^{\ast}(0)\right]d\hat{\Phi} = -\sqrt{\frac{(D-2)(D+1)}{2(D-1)}}
 \frac{\pi \alpha_s}{(D-1)(N_c^2-1)m_c^5}\delta(1-z),
\end{align}
\end{subequations}
The NLO QCD corrections to NRQCD LDMEs in production can be evaluated in a similar 
way as decay cases~\cite{He:2009bf,Li:2012rn}. After operator renormalization in $\overline{MS}$
scheme, we get
\begin{subequations}\label{swave_ldme}
\begin{align}
\langle0|\mathcal{O}^{c\bar{c}}(^3S_1^{[8]})|0\rangle|^{\mathcal{O}(\alpha_s)}_{\mathrm{NRQCD}}=
\frac{4\alpha_s}{3 \pi m_c^2}
\left( \frac{\mu_r^2}{\mu_\Lambda^2} \right)^{\epsilon}\left(-\frac{1}{\epsilon}-\ln 4\pi+\gamma_E\right)
\nonumber\\
\Big[C_F\langle0|\mathcal{O}(^3P_J^{[1]})|0\rangle|^{\mathrm{LO}}_{\mathrm{NRQCD}}+
B_F\langle0|\mathcal{O}(^3P_J^{[8]})|0\rangle|^{\mathrm{LO}}_{\mathrm{NRQCD}}\Big]
\end{align}
\begin{align}	
	\langle0|\mathcal{O}^{c\bar{c}}(^3S_1^{[8]})|0\rangle|^{\mathcal{O}(\alpha_sv^2)}_{\mathrm{NRQCD}}=
	-\frac{4\alpha_s}{5 \pi m_c^4}
	\left( \frac{\mu_r^2}{\mu_\Lambda^2} \right)^{\epsilon}\left(-\frac{1}{\epsilon}-\ln 4\pi+\gamma_E\right)
	\nonumber\\
\Big[C_F\langle0|\mathcal{P}(^3P_J^{[1]})|0\rangle|^{\mathrm{LO}}_{\mathrm{NRQCD}}
+B_F\langle0|\mathcal{P}(^3P_J^{[8]})|0\rangle|^{\mathrm{LO}}_{\mathrm{NRQCD}}\Big]
\end{align}
\begin{align}
	\langle0|\mathcal{P}^{c\bar{c}}(^3S_1^{[8]})|0\rangle|^{\mathcal{O}(\alpha_s)}_{\mathrm{NRQCD}}=
	\frac{4\alpha_s}{3 \pi m_c^2}
	\left( \frac{\mu_r^2}{\mu_\Lambda^2} \right)^{\epsilon}\left(-\frac{1}{\epsilon}-\ln 4\pi+\gamma_E\right)
	\nonumber\\
	\Big[C_F\langle0|\mathcal{P}(^3P_J^{[1]})|0\rangle|^{\mathrm{LO}}_{\mathrm{NRQCD}}+
	B_F\langle0|\mathcal{P}(^3P_J^{[8]})|0\rangle|^{\mathrm{LO}}_{\mathrm{NRQCD}}\Big]
\end{align}
\begin{align}
	\langle0|\mathcal{P}^{c\bar{c}}(^3S_1^{[8]},^3D_1^{[8]})|0\rangle|^{\mathcal{O}(\alpha_s)}_{\mathrm{NRQCD}}=
	\frac{4\alpha_s}{3 \pi m_c^2}
	\left( \frac{\mu_r^2}{\mu_\Lambda^2} \right)^{\epsilon}\left(-\frac{1}{\epsilon}-\ln 4\pi+\gamma_E\right)
	\nonumber\\
	G_{J}\Big[C_F\langle0|\mathcal{P}(^3P_J^{[1]})|0\rangle|^{\mathrm{LO}}_{\mathrm{NRQCD}}+
	B_F\langle0|\mathcal{P}(^3P_J^{[8]})|0\rangle|^{\mathrm{LO}}_{\mathrm{NRQCD}}\Big]
\end{align}
\end{subequations}
where $C_F=(N_c^2-1)/2/N_c$, $B_F=(N_c^2-4)/4/N_c$, and the coefficients $G_J$ are the generalized 
Clebsch-Gordan coefficients in $D$-dimension given by
\begin{subequations}
	\begin{align}
		G_0 = & -\sqrt{\frac{2(D-2)}{(D-1)(D+1)}} \, , \\
		G_1 = & \sqrt{\frac{2}{(D-2)(D-1)(D+1)}} \, , \\
		G_2 = & -\frac{D-3}{D+1} \sqrt{\frac{2}{(D-2)(D-1)(D+1)}} \, .
	\end{align}	
\end{subequations}

Plug the SDCs in Eq.[\ref{swave_sdc}] and LDMEs in Eq.[\ref{swave_ldme}] into 
Eq.[\ref{nrqcd_p_ff}], match it and with the results in Eq.[\ref{qcd_p_ff}], we then get 
the finite of results of the SDCs for $g\to c\bar{c}(^3P_J^{[1,8]})$. Our $v^2$ LO results 
agree with those in literature and the $v^2$ corrections are new. For completeness, here 
we list all of them below:

\begin{subequations}\label{eq:sdc}
	\begin{align}
		d^{(0)}[g\to c\bar{c}(^3P_0^{[1]})] &=\frac{ \alpha_s^2 }{ N_c m_c^5 }
		\Bigg[  \left(\frac{1}{9} - \frac{2}{9} \ln \left(\frac{\mu_\Lambda ^2}{4 m_c^2} \right) \right) \delta(1-z) \nonumber \\
		&+\frac{4}{9[1-z]_+}
		+\frac{-26 z^2+85 z-8}{18}+ (5-3 z) \ln(1-z) \Bigg] \,,
	\end{align}
	\begin{align}
		d^{(2)}[g\to c\bar{c}(^3P_0^{[1]})] &=\frac{ \alpha_s^2 }{ N_c m_c^5 }
		\Bigg[  \left(\frac{31}{45} \ln \left(\frac{\mu_\Lambda ^2}{4 m_c^2} \right) -\frac{71}{450}  \right) \delta(1-z) \nonumber \\
		&-\frac{62}{45[1-z]_+}
		+\frac{710 z^2-2419 z+248}{180} + \frac{255 z-433}{30} \ln(1-z) \Bigg] \,, 
	\end{align}
	\begin{align}
		d^{(0)}[g\to c\bar{c}(^3P_1^{[1]})] =\frac{ \alpha_s^2 }{ N_c m_c^5 }
		\Bigg[  \left(\frac{1}{18} - \frac{2}{9} \ln 
		\left(\frac{\mu_\Lambda ^2}{4 m_c^2} \right) \right) 
		\delta(1-z)+\frac{4}{9[1-z]_+}+\frac{-4 z^2-z-4}{9} \Bigg]  \,,
	\end{align}
	\begin{align}
		d^{(2)}[g\to c\bar{c}(^3P_1^{[1]})] =\frac{ \alpha_s^2 }{ N_c m_c^5 }
		\Bigg[  \left(\frac{7}{15} \ln \left(\frac{\mu_\Lambda ^2}{4 m_c^2} \right) + \frac{7}{100} \right) \delta(1-z) -\frac{14}{15[1-z]_+}
		+\frac{7(4 z^2+z+4)}{30} \Bigg]  \,, 
	\end{align}
	\begin{align}
		d^{(0)}[g\to c\bar{c}(^3P_2^{[1]})]  &=\frac{ \alpha_s^2 }{ N_c m_c^5 }
		\Bigg[  \left(\frac{7}{90} - \frac{2}{9} \ln \left(\frac{\mu_\Lambda ^2}{4 m_c^2} \right) \right) \delta(1-z) \nonumber \\
		&+\frac{4}{9[1-z]_+}
		+\frac{-4 z^2+11 z-4}{9}-\frac{4(z-2)}{5} \ln(1-z) \Bigg] \,,
	\end{align}
	\begin{align}
		d^{(2)}[g\to c\bar{c}(^3P_2^{[1]})] &=\frac{ \alpha_s^2 }{ N_c m_c^5 }
		\Bigg[  \left(\frac{5}{9} \ln \left(\frac{\mu_\Lambda ^2}{4 m_c^2} \right) -\frac{7}{900} \right) \delta(1-z) \nonumber \\
		&-\frac{10}{9[1-z]_+}
		+\frac{476 z^2-1375 z+500}{450} +\frac{2(75 z-152)}{75} \ln(1-z) \Bigg] \,, 
	\end{align}
	\begin{align}
	d^{(0,2)}[g\to c\bar{c}(^3P_J^{[8]})]=\frac{B_F}{C_F}d^{(0,2)}[g\to c\bar{c}(^3P_J^{[1]})]
	\end{align}
\end{subequations}

Our results show that unlike the $S$-wave cases~\cite{Bodwin:2003wh,Gao:2016ihc}, the SDCs 
of the $v^2$ corrections to FF are not proportionals to those at $v^2$ LO.

\section{Numerical results and discussion} \label{sec:numerical}
\begin{figure*}[tb!]
	\centering
	\begin{minipage}{0.32\textwidth}
		\centerline{\includegraphics[width=1\textwidth]{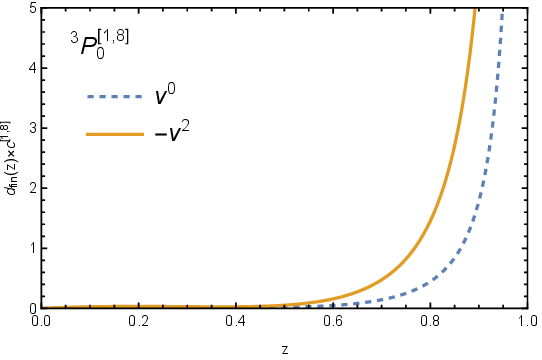}}
	\end{minipage}
	\begin{minipage}{0.32\textwidth}
		\centerline{\includegraphics[width=1\textwidth]{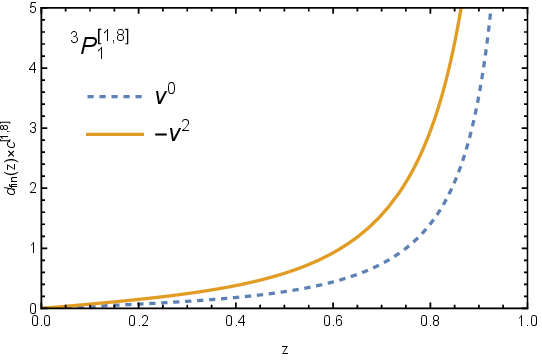}}
	\end{minipage}
	\begin{minipage}{0.32\textwidth}
		\centerline{\includegraphics[width=1\textwidth]{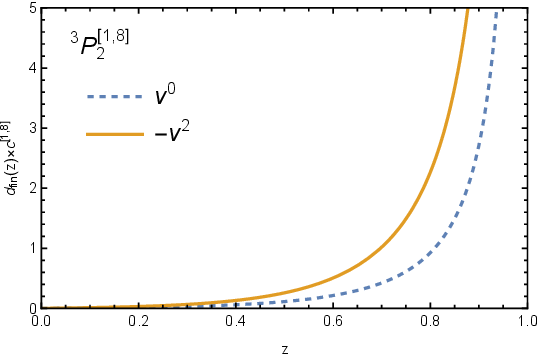}}
	\end{minipage}
	\caption{The finite part of the SDCs of the fragmentation functions of $g\to c\bar{c} (^3P_J^{[1,8]})$ for $J=0$(left), $J=1$(middle), and $J=2$(right) at QCD LO. The dashed line is for $d_{\mathrm{fin}}^{(0)}(z)\times c^{[1,8]}$, and the solid line is for $d_{\mathrm{fin}}^{(2)}(z)) \times c^{[1,8]}$, where $c^{[1]}=3m_c^5/\alpha_s^2$ for color-singlet and  $c^{[8]}=15m_c^5/(16\alpha_s^2)$ for color-octet. \label{fig:sdcs}}
\end{figure*}

We are now in position to present the numerical investigation of the $v^2$ corrections. We 
first show the behaviors of the finite part of the SDCs, denoted as 
$d_{\mathrm{fin}}^{(0)}[g\to c\bar{c}(^3P_J^{[1,8]})]$ and 
$d_{\mathrm{fin}}^{(2)}[g\to c\bar{c}(^3P_J^{[1,8]})]$, in which the singular terms at 
$z=1$ from the $\delta$ and ``+" functions in Eq.~\ref{eq:sdc} are excluded, as function 
of the momentum fraction $z$ in Fig.~\ref{fig:sdcs}. The overall scaling factors are 
defined as $c^{[1]}=3m_c^5/\alpha_s^2$ for CS and 
$c^{[8]}=15m_c^5/(16\alpha_s^2)$ for CO. As shown in Fig.~\ref{fig:sdcs}, 
$d_{\mathrm{fin}}^{(0)}[g\to c\bar{c}(^3P_J^{[1,8]})]$ are positive for all $J=0,1,2$,  
channels, while the $d_{\mathrm{fin}}^{(2)}[g\to c\bar{c}(^3P_J^{[1,8]})]$ are all 
negative conversely. Besides, their absolute values exhibit similar dependence on $z$, 
with a notable increase in magnitude as $z$ approaches 1 indicating a significant 
contribution from the large $z$ region. 

\begin{figure}[!htp]
	\centering
	\includegraphics[width=0.45\textwidth]{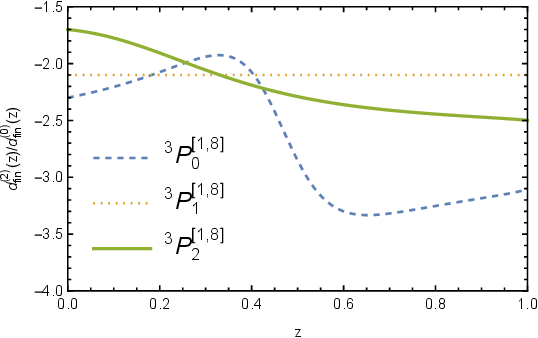}
	\caption{ The ratios between the finite part of the SDCs, 
		$d_{\mathrm{fin}}^{(2)}[g\to c\bar{c}(^3P_J^{[1,8]})]$ and 
		$d_{\mathrm{fin}}^{(0)}[g\to c\bar{c}(^3P_J^{[1,8]})]$, 
		for $J=0,1,2$, in which the $\delta$ and ``+" function terms are dropped.}
	\label{fig:sdcratio}
\end{figure}

The plots in Fig.~[\ref{fig:sdcs}] also tell clearly that 
$d_{\mathrm{fin}}^{(2)}[g\to c\bar{c}(^3P_J^{[1,8]})]$ are much larger than 
$d_{\mathrm{fin}}^{(0)}[g\to c\bar{c}(^3P_J^{[1,8]})]$ in magnitude, which suggests that 
the relativistic corrections can not be ignored in gluon fragmentation process. To know 
more clearly about the relations between $v^2$ LO and sub-LO SDCs of gluon fragmentation 
into $c\bar{c}(^3P_J^{[1,8]})$ channels, we also calculate the ratios 
between $d^{(2)}_{\mathrm{fin}}[g\to c\bar{c}(^3P_J^{[1,8]})]$ and 
$d^{(0)}_{\mathrm{fin}}[g\to c\bar{c}(^3P_J^{[1,8]})]$, for $J=0,1,2$, and the results are 
shown in Fig.[\ref{fig:sdcratio}]. We find that for $J=1$ channel the ratio is a constant 
of $-21/10$. In contrast, for the other $J=0$ and $2$ channels the ratios exhibit an 
obvious dependence on $z$. These behaviors are different from the previous results for 
$S$-wave cases that observed in Ref.~\cite{Bodwin:2003wh,Gao:2016ihc}, where the SDC of 
$v^2$ correction part is simply a product of the $v^2$ LO SDC and a constant $-11/6$.

Next, we will investigate the phenomenological influence of the complete relativistic 
corrections in which the contribution of $\delta$ and ``+" function terms in 
Eqs.~\ref{eq:sdc} will be taken into account too. We consider the CMS measurements at 
the $\sqrt{7}$ TeV LHC in rapidity range of $|y|<1.2$ as an example, in which the $p_T$ 
of $J/\psi$ can reach to $120$ GeV~\cite{CMS:2015lbl}. According to the factorization 
formula proposed in Ref.~\cite{Kang:2011zza,Kang:2011mg,Kang:2014tta}, the differential 
cross section from gluon fragmentation contribution into heavy quarkonium $H$ 
hadroproduction in large $p_T$ region can be expressed as
\begin{align}\label{eq:collins}
E_p\frac{d\sigma_{A+B\to H+X}}{d^3p} \approx \int \frac{dz}{z^2}D_{g\to H}(z;\mu)
E_c\frac{\hat{\sigma} _{A+B \rightarrow g(p_g)+X'}}{d^3p_c} (p_g=p/z)
\end{align}
Combined it with NRQCD factorization of FF, we get up to $v^2$ sub-LO order, 
\begin{align}
\frac{d\sigma}{dp_T}= \frac{\ud F (n)}{\ud p_T} \langle \mathcal{O}^{H}(n)\rangle+ \frac{\ud G (n)}{d p_T} \frac{\langle \mathcal{P}^{H}(n)\rangle}{m_c^2} \,
\end{align}
where $F(n)$ and $G(n)$ are the convolution of gluon production differential cross section 
with the SDCs of its FF. 

It was found that for $n=$$^3P_J^{[8]}$ production the LP contribution in 
Eq.[\ref{eq:collins}] can well reproduce the exact NLO QCD results by convoluting the FF 
ignoring its evolution with gluon partonic cross section calculated at QCD LO and the NLO 
PDF set ~\cite{Ma:2014svb}. Specifically, they take $m_c=1.5$ GeV, set the normalization 
and factorization scales $\mu_r=\mu_f=p_T$ and the NRQCD factorization scale 
$\mu_\Lambda = m_c$, use one-loop running of $\alpha_s$ with $n_f=5$ flavors active 
quark and $\Lambda_{QCD}^{(5)}=165~\mathrm{GeV}$, and choose the CTEQ6M set for parton 
distribution functions (PDFs) \cite{Pumplin:2002vw}. For the same token, we adopt the 
same parameter settings.

\begin{figure*}[!tb]
\centering
\includegraphics[width=0.45\textwidth]{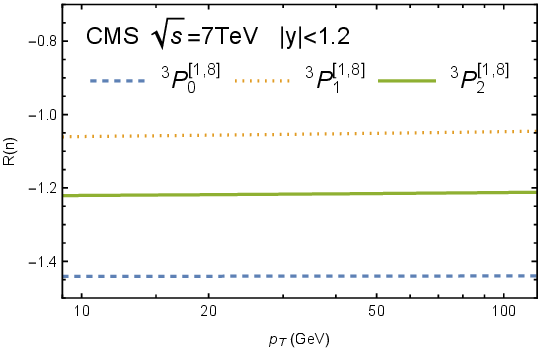}
\caption{ The ratios $R(n)$ for $n=$$^3P_{J}^{[1,8]}$ as function of $p_T$ under CMS conditions at the $\sqrt{s}=7$ TeV LHC in rapidity range of $|y|<1.2$.}
\label{fig:csratio0}
\end{figure*}

The significance of relativistic corrections then can be seen easily from the ratio 
\begin{align}
R (n) &= \frac{\ud G (n)}{\ud p_T} \Bigg/ \frac{\ud F (n)}{\ud p_T} \, ,
\end{align}
The results of $R(n)$ for $n=$$^3P_{J}^{[1,8]}$ with $J=0,1,2$ are shown separately in 
Fig.[\ref{fig:csratio0}]. We observe unlike the ratios between SDCs of FF the values of 
$|R(n)|$ are most constant of $\mathcal{O}(1)$, which implies that the effect of 
relativistic corrections are also of $\mathcal{O}(v^2)$ given the ratio between NRQCD 
LDMEs $\langle v_n^{2}\rangle^{H}=\frac{\langle \mathcal{P}^{H}(n)\rangle}
{m_c^2 \langle \mathcal{O}^{H}(n)\rangle}\sim\mathcal{O}(v^2)$ obeying the velocity 
scaling rule. 

\begin{figure*}[tb!]
\centering
\includegraphics[width=0.3\textwidth]{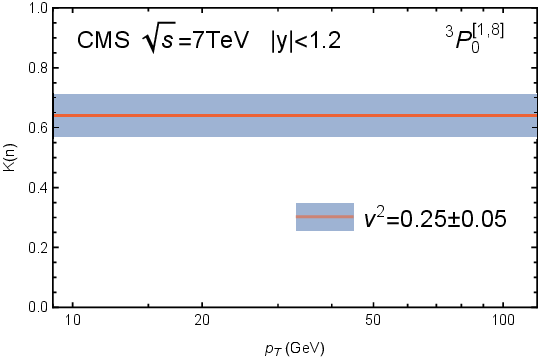}
\includegraphics[width=0.3\textwidth]{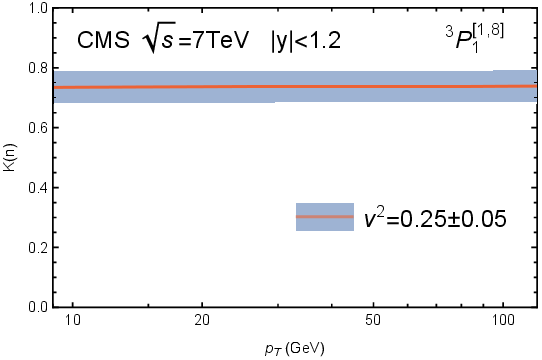}
\includegraphics[width=0.3\textwidth]{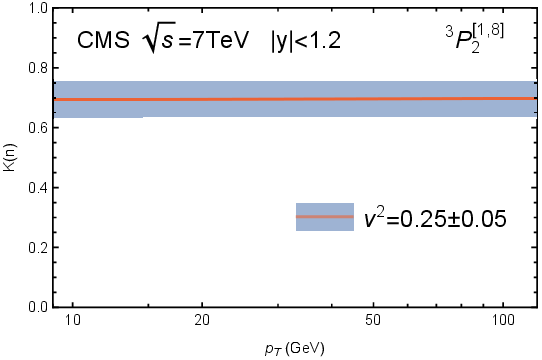}
\caption{ The K-factors $K(n)$ for $n=$$^3P_{J}^{[1,8]}$ as function of $p_T$ under CMS 
conditions at the $\sqrt{s}=7$ TeV LHC in rapidity range of $|y|<1.2$. The central solid 
line is calculated at $\langle v_n^{2}\rangle^{H}=0.25$, and  the band is obtained by 
varying $\langle v_n^{2}\rangle^{H}$ from 0.2 to 0.3.}\label{fig:csratio}
\end{figure*}

To quantify the effects of relativistic corrections on the differential cross sections 
for gluon fragmentation into $c\bar{c}(^3P_J^{[1,8]})$, similar to NLO QCD corrections we 
introduce the $K-$factor as the ratios between $v^2$ NLO and LO predictions, which can be 
expressed in a neat form as 
\begin{eqnarray}
	K(n)=1+ R (n) \langle v_n^{2}\rangle^{H} 
\end{eqnarray} 
In Ref.~\cite{Bodwin:2003wh}, $\langle v_n^2\rangle^{J/\psi}$ was roughly 
estimated to be 0.21 for both CS and CO states by using the naive Gremm-Kapustin 
relation~\cite{Gremm:1997dq}, and potential model calculation yield that for CS S-wave 
case $\langle v_{^3S_1^{[1]}}^{2}\rangle^{J/\psi}=0.25\pm0.05\pm0.08$~\cite{Bodwin:2006dn} 
or $\langle v_{^3S_1^{[1]}}^{2}\rangle^{J/\psi}=0.225^{+0.106}_{-0.088}$~\cite{Bodwin:2007fz}, 
which are compatible with each other. In our numerical analysis, we take 
$\langle v_n^{H}\rangle=0.25 \pm 0.05$ for $n=$$^3P_J^{[8]}$ of $H=J/\psi$ and 
$n=$$^3P_J^{[1]}$ of $H=\chi_{cJ}$ as a rough estimation. The results of $K(n)$ are shown 
in Fig.[\ref{fig:csratio}]. As can be seen directly, the relativistic corrections 
significantly diminish the LO predictions for all channels in the whole $p_T$ range.
Specifically, when $\langle v_n^{2}\rangle^{H}=0.25$, the relativistic corrections lead to a reduction of $36\%,26\%$ and $30\%$ for $J=0,1,2$, respectively. 
Moreover, the impact of relativistic corrections become more pronounced as $v^2$ 
increases. When $\langle v_n^{2}\rangle^{H}$ goes from $0.2$ to $0.3$, the reduction in 
differential cross section escalates from approximately $29\%$ to $43\%$ for $\chi_{c0}$, 
from $21\%$ to $32\%$ for $\chi_{c1}$, and from $24\%$ to $36\%$ for $\chi_{c2}$. 
In addition, we also checked that the $K(n)$-factors the same as $R(n)$ almost do not 
depend on the rapidity cut. 

In our previous work, we have already pointed out that higher order relativistic effects 
are considerable in moderate $p_T$ region of $J/\psi$ photo- and 
hadro-production~\cite{He:2014sga}. In line with that, through calculation of the 
relativistic corrections to the gluon FFs, we affirm that in large $p_T$ region the higher 
order relativistic effects are important as well. Currently, the all the CO NRQCD LDMEs 
for $J/\psi$ production were extracted through fitting to the experimental data with only 
the NLO QCD corrections. It was found that there is a large cancellation between 
$\langle\mathcal{O}(^3P_J^{[8]})\rangle^{J/\psi}$ and 
$\langle\mathcal{O}(^3S_1^{[8]})\rangle^{J/\psi}$~\cite{Ma:2010yw}. Such a small value 
also results in un-polarized $J/\psi$ hadroproduction in large $p_T$ 
region~\cite{Shao:2014yta}. We think it will be very imperative to incorporate the 
relativistic corrections when to fit the NRQCD LDMEs in the future work. 

\section{Conclusions}

Within NRQCD factorization, the gluon FFs into $^3P_J^{[1,8]}$ are calculated up to 
the $v^2$-sub LO. The infrared finite SDCs are obtained via the matching procedure that 
equates the results of full QCD and NRQCD computations. In full QCD calculation, the 
spinor projection method is implemented.  In NRQCD calculation the $S-D$ mixing 
contribution should be included which is necessary to remove the infrared divergences in the full QCD results. In the meantime the NRQCD LDME 
$\langle0|\mathcal{O}^{c\bar{c}}(^3S_1^{[8]})|0\rangle$ is expanded up to 
$\mathcal{O}(\alpha_{s}v^2)$. Different from $S$-wave case, the SDCs of $^3P_J^{[1,8]}$ 
at the $v^2$-sub LO petition a new function of the momentum fraction $z$. We then 
convolute the FFs with gluon hadroproduction cross section at QCD LO and find that the 
effects of NLO relativistic corrections are almost not affected by $p_T$ or rapidity of 
charmonium state and numerically the relativistic corrections can lead to about $30\%$ 
reduction to the LO predictions. Therefore they can not be ignored when investigating the 
$J/\psi$ production mechanism. We then conclude that to finally reveal the production 
mechanism of heavy quarkonium, the relativistic corrections are indispensable.

\begin{acknowledgments}
This work was supported in part by the German Research Foundation DFG through Grants No. 
KN 365/13-2 and 365/14-2, by the National Natural Science Foundation of China through 
Grants No. 12205010, and by Fundamental Research Funds for the Central Universities 
through Grant No. buctrc202432.

\end{acknowledgments}



\begin{thebibliography}{99}
\bibitem{Bodwin:1994jh}
G.~T.~Bodwin, E.~Braaten and G.~P.~Lepage,
``Rigorous QCD analysis of inclusive annihilation and production of heavy quarkonium,''
Phys. Rev. D \textbf{51} (1995), 1125-1171
[erratum: Phys. Rev. D \textbf{55} (1997), 5853].

\bibitem{Caswell:1985ui}
W.~E.~Caswell and G.~P.~Lepage,
``Effective Lagrangians for Bound State Problems in QED, QCD, and Other Field Theories,''
Phys. Lett. B \textbf{167} (1986), 437-442.

\bibitem{Lansberg:2019adr}
J.~P.~Lansberg,
``New Observables in Inclusive Production of Quarkonia,''
Phys. Rept. \textbf{889} (2020), 1-106.


\bibitem{Ma:2010yw}
Y.~Q.~Ma, K.~Wang and K.~T.~Chao,
``$J/\psi (\psi^\prime)$ production at the Tevatron and LHC at ${\cal O}(\alpha_s^4v^4)$ in nonrelativistic QCD,''
Phys. Rev. Lett. \textbf{106} (2011), 042002.

\bibitem{Butenschoen:2011yh}
M.~Butenschoen and B.~A.~Kniehl,
``World data of $J/\psi$ production consolidate NRQCD factorization at NLO,''
Phys. Rev. D \textbf{84} (2011), 051501.

\bibitem{Gong:2012ug}
B.~Gong, L.~P.~Wan, J.~X.~Wang and H.~F.~Zhang,
``Polarization for Prompt J/{\ensuremath{\psi}} and {\ensuremath{\psi}}(2s) Production at the Tevatron and LHC,''
Phys. Rev. Lett. \textbf{110} (2013) no.4, 042002.

\bibitem{Bodwin:2014gia}
G.~T.~Bodwin, H.~S.~Chung, U.~R.~Kim and J.~Lee,
``Fragmentation contributions to $J/\psi$ production at the Tevatron and the LHC,''
Phys. Rev. Lett. \textbf{113} (2014) no.2, 022001.

\bibitem{Brambilla:2024iqg}
N.~Brambilla, M.~Butenschoen and X.~P.~Wang,
``How well does nonrelativistic QCD factorization work at next-to-leading order?,''
Phys. Rev. D \textbf{112} (2025) no.1, 1.
	
\bibitem{Braaten:1993rw}
E.~Braaten and T.~C.~Yuan,
``Gluon fragmentation into heavy quarkonium,''
Phys. Rev. Lett. \textbf{71} (1993), 1673-1676.
	
\bibitem{Braaten:1993mp}
E.~Braaten, K.~m.~Cheung and T.~C.~Yuan,
``$Z_0$ decay into charmonium via charm quark fragmentation,''
Phys. Rev. D \textbf{48} (1993), 4230-4235.

\bibitem{Kang:2011zza}
Z.~B.~Kang, J.~W.~Qiu and G.~Sterman,
``Factorization and quarkonium production,''
Nucl. Phys. B Proc. Suppl. \textbf{214} (2011), 39-43.

\bibitem{Kang:2011mg}
Z.~B.~Kang, J.~W.~Qiu and G.~Sterman,
``Heavy quarkonium production and polarization,''
Phys. Rev. Lett. \textbf{108} (2012), 102002.

\bibitem{Kang:2014tta}
Z.~B.~Kang, Y.~Q.~Ma, J.~W.~Qiu and G.~Sterman,
``Heavy Quarkonium Production at Collider Energies: Factorization and Evolution,''
Phys. Rev. D \textbf{90} (2014) no.3, 034006.

\bibitem{Braaten:1994kd}
E.~Braaten and T.~C.~Yuan,
``Gluon fragmentation into P wave heavy quarkonium,''
Phys. Rev. D \textbf{50} (1994), 3176-3180.

\bibitem{Cho:1994gb}
P.~L.~Cho, M.~B.~Wise and S.~P.~Trivedi,
``Gluon fragmentation into polarized charmonium,''
Phys. Rev. D \textbf{51} (1995), R2039-R2043.

\bibitem{Ma:1995vi}
J.~P.~Ma,
``Quark fragmentation into p wave triplet quarkonium,''
Phys. Rev. D \textbf{53} (1996), 1185-1190.

\bibitem{Braaten:1995cj}
E.~Braaten and T.~C.~Yuan,
``Gluon fragmentation into spin triplet S wave quarkonium,''
Phys. Rev. D \textbf{52} (1995), 6627-6629.

\bibitem{Beneke:1995yb}
M.~Beneke and I.~Z.~Rothstein,
``Psi-prime polarization as a test of color octet quarkonium production,''
Phys. Lett. B \textbf{372} (1996), 157-164
[erratum: Phys. Lett. B \textbf{389} (1996), 769].

\bibitem{Braaten:1996rp}
E.~Braaten and Y.~Q.~Chen,
``Dimensional regularization in quarkonium calculations,''
Phys. Rev. D \textbf{55} (1997), 2693-2707.

\bibitem{Ma:2013yla}
Y.~Q.~Ma, J.~W.~Qiu and H.~Zhang,
``Heavy quarkonium fragmentation functions from a heavy quark pair. I. $S$ wave,''
Phys. Rev. D \textbf{89} (2014) no.9, 094029.

\bibitem{Ma:2015yka}
Y.~Q.~Ma, J.~W.~Qiu and H.~Zhang,
``Fragmentation functions of polarized heavy quarkonium,''
JHEP \textbf{06} (2015), 021.

\bibitem{Zhang:2017xoj}
P.~Zhang, Y.~Q.~Ma, Q.~Chen and K.~T.~Chao,
``Analytical calculation for the gluon fragmentation into spin-triplet S-wave quarkonium,''
Phys. Rev. D \textbf{96} (2017) no.9, 094016.

\bibitem{Braaten:2000pc}
E.~Braaten and J.~Lee,
``Next-to-leading order calculation of the color octet 3S(1) gluon fragmentation function for heavy quarkonium,''
Nucl. Phys. B \textbf{586} (2000), 427-439.

\bibitem{Artoisenet:2018dbs}
P.~Artoisenet and E.~Braaten,
``Gluon fragmentation into quarkonium at next-to-leading order using FKS subtraction,''
JHEP \textbf{01} (2019), 227.

\bibitem{Feng:2018ulg}
F.~Feng and Y.~Jia,
``Next-to-leading-order QCD corrections to gluon fragmentation into quarkonia*,''
Chin. Phys. C \textbf{47} (2023) no.3, 033103.

\bibitem{Zhang:2018mlo}
P.~Zhang, C.~Y.~Wang, X.~Liu, Y.~Q.~Ma, C.~Meng and K.~T.~Chao,
``Semi-analytical calculation of gluon fragmentation into$^{1}$S$_{0}^{[1,8]}$ quarkonia at next-to-leading order,''
JHEP \textbf{04} (2019), 116.

\bibitem{Zhang:2020atv}
P.~Zhang, C.~Meng, Y.~Q.~Ma and K.~T.~Chao,
``Gluon fragmentation into $^{3} {P}_J^{\left[1,8\right]} $ quark pair and test of NRQCD factorization at two-loop level,''
JHEP \textbf{08} (2021), 111.

\bibitem{Zheng:2019dfk}
X.~C.~Zheng, C.~H.~Chang and X.~G.~Wu,
``NLO fragmentation functions of heavy quarks into heavy quarkonia,''
Phys. Rev. D \textbf{100} (2019) no.1, 014005.

\bibitem{Sepahvand:2017gup}
R.~Sepahvand and S.~Dadfar,
``NLO corrections to $c$- and $b$-quark fragmentation into $J/\psi$ and $\Upsilon$,''
Phys. Rev. D \textbf{95} (2017) no.3, 034012,
doi:10.1103/PhysRevD.95.034012.

\bibitem{Feng:2021uct}
F.~Feng, Y.~Jia and W.~L.~Sang,
``Next-to-leading-order QCD corrections to heavy quark fragmentation into ${}^1S^{(1,8)}_0$ quarkonia,''
Eur. Phys. J. C \textbf{81} (2021) no.7, 597.

\bibitem{He:2007te}
Z.-G.~He, Y.~Fan, and K.-T.~Chao,
``Relativistic corrections to $J/\psi$ exclusive and inclusive double charm production at $B$ factories,''
Phys.\ Rev.\ D \textbf{75} (2007) 074011.

\bibitem{He:2009uf}
Z.-G.~He, Y.~Fan, and K.-T.~Chao,
``Relativistic correction to $e^+e^-\to J/\psi + gg$ at $B$ factories and constraint on color-octet matrix elements,''
Phys.\ Rev.\ D \textbf{81} (2010) 054036.

\bibitem{Jia:2009np}
Y.~Jia,
``Color-singlet relativistic correction to inclusive $J/\psi$ production associated with light hadrons at $B$ factories,''
Phys.\ Rev.\ D \textbf{82} (2010) 034017.

\bibitem{Xu:2012am}
G.-Z.~Xu, Y.-J.~Li, K.-Y.~Liu, and Y.-J.~Zhang,
``Relativistic correction to color-octet $J/\psi$ production at hadron colliders,''
Phys.\ Rev.\ D \textbf{86} (2012) 094017.

\bibitem{He:2014sga}
Z.~G.~He and B.~A.~Kniehl,
``Relativistic corrections to prompt $J/\psi$ photo- and hadroproduction,''
Phys.\ Rev.\ D \textbf{90} (2014) 014045
[erratum: Phys.\ Rev.\ D \textbf{94} (2016) 079903].

\bibitem{He:2015gla}
Z.-G.~He and B.~A.~Kniehl,
``Relativistic corrections to $J/\psi$ polarization in photo- and hadroproduction,''
Phys.\ Rev.\ D \textbf{92} (2015) 014009.



\bibitem{He:2024ugx}
Z.~G.~He, X.~B.~Jin and B.~A.~Kniehl,
``Relativistic corrections to prompt double charmonium hadroproduction near threshold,''
Phys. Rev. D \textbf{109} (2024) no.9, 094013.

\bibitem{Bodwin:2003wh}
G.~T.~Bodwin and J.~Lee,
``Relativistic corrections to gluon fragmentation into spin triplet S wave quarkonium,''
Phys. Rev. D \textbf{69} (2004), 054003.

\bibitem{Bodwin:2012xc}
G.~T.~Bodwin, U.~R.~Kim and J.~Lee,
``Higher-order relativistic corrections to gluon fragmentation into spin-triplet S-wave quarkonium,''
JHEP \textbf{11} (2012), 020
[erratum: JHEP \textbf{07} (2023), 170].

\bibitem{Sang:2009zz}
W.~l.~Sang, L.~f.~Yang and Y.~q.~Chen,
``Relativistic corrections to heavy quark fragmentation to S-wave heavy mesons,''
Phys. Rev. D \textbf{80} (2009), 014013.

\bibitem{Cui:2025wjq}
S.~Cui, Y.~J.~Li, G.~Z.~Xu and K.~Y.~Liu,
``Order-$v^4$ corrections to heavy quark fragmentation to S-wave heavy quarkonium,''
[arXiv:2512.20539 [hep-ph]].

\bibitem{Gao:2016ihc}
X.~Gao, Y.~Jia, L.~Li and X.~Xiong,
``Relativistic correction to gluon fragmentation function into pseudoscalar quarkonium,''
Chin. Phys. C \textbf{41} (2017) no.2, 023103.

\bibitem{Collins:1981uw}
J.~C.~Collins and D.~E.~Soper,
``Parton Distribution and Decay Functions,''
Nucl. Phys. B \textbf{194} (1982), 445-492.

\bibitem{Petrelli:1997ge}
A.~Petrelli, M.~Cacciari, M.~Greco, F.~Maltoni and M.~L.~Mangano,
``NLO production and decay of quarkonium,''
Nucl. Phys. B \textbf{514} (1998), 245-309

\bibitem{Bodwin:2002cfe}
G.~T.~Bodwin and A.~Petrelli,
``Order-$v^4$ corrections to $S$-wave quarkonium decay,''
Phys. Rev. D \textbf{66} (2002), 094011
[erratum: Phys. Rev. D \textbf{87} (2013) no.3, 039902]

\bibitem{He:2009bf}
Z.~G.~He, Y.~Fan and K.~T.~Chao,
``NRQCD Predictions of D-Wave Quarkonia D-3(J) (J = 1,2,3) Decay into Light Hadrons at Order alpha**3 (S),''
Phys. Rev. D \textbf{81} (2010), 074032.

\bibitem{Li:2012rn}
J.~Z.~Li, Y.~Q.~Ma and K.~T.~Chao,
``QCD and Relativistic $O(\alpha_{s}v^2)$ Corrections to Hadronic Decays of Spin-Singlet Heavy Quarkonia $h_c, h_b$ and $\eta_b$,''
Phys. Rev. D \textbf{88} (2013) no.3, 034002.

\bibitem{CMS:2015lbl}
V.~Khachatryan \textit{et al.} [CMS],
``Measurement of J/{\ensuremath{\psi}} and {\ensuremath{\psi}}(2S) Prompt Double-Differential Cross Sections in pp Collisions at $\sqrt{s}$=7  TeV,''
Phys. Rev. Lett. \textbf{114} (2015) no.19, 191802

\bibitem{Ma:2014svb}
Y.~Q.~Ma, J.~W.~Qiu, G.~Sterman and H.~Zhang,
``Factorized power expansion for high-$p_T$ heavy quarkonium production,''
Phys. Rev. Lett. \textbf{113} (2014) no.14, 142002.

\bibitem{Pumplin:2002vw}
J.~Pumplin, D.~R.~Stump, J.~Huston, H.~L.~Lai, P.~M.~Nadolsky and W.~K.~Tung,
``New generation of parton distributions with uncertainties from global QCD analysis,''
JHEP \textbf{07} (2002), 012.

\bibitem{Gremm:1997dq}
M.~Gremm and A.~Kapustin,
``Annihilation of S wave quarkonia and the measurement of alpha-s,''
Phys. Lett. B \textbf{407} (1997), 323-330.

\bibitem{Bodwin:2006dn}
G.~T.~Bodwin, D.~Kang and J.~Lee,
``Potential-model calculation of an order-v(2) NRQCD matrix element,''
Phys. Rev. D \textbf{74} (2006), 014014.

\bibitem{Bodwin:2007fz}
G.~T.~Bodwin, H.~S.~Chung, D.~Kang, J.~Lee and C.~Yu,
``Improved determination of color-singlet nonrelativistic QCD matrix elements for S-wave charmonium,''
Phys. Rev. D \textbf{77} (2008), 094017.

\bibitem{Shao:2014yta}
H.~S.~Shao, H.~Han, Y.~Q.~Ma, C.~Meng, Y.~J.~Zhang and K.~T.~Chao,
``Yields and polarizations of prompt $J/\psi$ and $\psi(2S)$ production in hadronic collisions,''
JHEP \textbf{05} (2015), 103.
	
\end{thebibliography}

\end{document}